\documentclass[twocolumn]{article}

\usepackage[margin=.5in]{geometry}
\usepackage{mathpazo}
\usepackage{makecell}
\usepackage{bm}

\usepackage{float}
\usepackage{lipsum}

\usepackage{comment}
\usepackage{makecell}
\usepackage{colortbl}
\usepackage[utf8]{inputenc}
\usepackage{microtype}
\usepackage{xcolor,pifont}
\usepackage[T1]{fontenc}

\usepackage{tikz}
\usepackage{pgfplots}
\pgfplotsset{compat=1.18}

\usepackage{microtype}
\usepackage{graphicx}
\usepackage{booktabs} 
\usepackage[hidelinks]{hyperref}

\usepackage{amsmath}
\usepackage{amssymb}
\usepackage{mathtools}
\usepackage{amsthm}

\usepackage{algorithmic}
\usepackage{xspace}
\usepackage{subcaption}

\usepackage{bm}

\theoremstyle{plain}

\theoremstyle{definition}

\theoremstyle{remark}

\usepackage{xcolor}
\usepackage{soul}

\usepackage{color}
\definecolor{light}{rgb}{0.5, 0.5, 0.5}

\usepackage{xcolor}
\usepackage{soul}
\usepackage[normalem]{ulem}

\newif\ifdraft\drafttrue
\newif\ifaftersubmission\aftersubmissionfalse
\newif\iflater\latertrue 
\newif\ifjustbeforesubmission\justbeforesubmissionfalse  
\newif\ifextended\extendedfalse

\definecolor{dkred}{rgb}{0.7,0,0}
\definecolor{dkpurple}{HTML}{4e02eb}
\definecolor{dkgreen}{HTML}{006329}
\definecolor{dkblue}{HTML}{2d5491}
\definecolor{dkorange}{HTML}{825a23}
\definecolor{ltgreen}{HTML}{3a9e67}
\definecolor{teal}{HTML}{007982}
\definecolor{fuchsia}{HTML}{8C368C}

\usepackage[addedmarkup=uline, defaultcolor=magenta, todonotes={textsize=scriptsize, textwidth=2.2cm}, authormarkuptext=name, commandnameprefix=always, xcolor]{changes} 

\definechangesauthor[name={CW}, color=dkgreen]{cw}

\title{\textbf{F2: Offline Reinforcement Learning for Hamiltonian Simulation via \emph{\uline{F}ree-\uline{F}ermionic} Subroutine Compilation}}

\author{
    Ethan Decker\footnote{Corresponding author. \texttt{ecd5249@upenn.edu}} \\
    University of Pennsylvania \& Pacific Northwest National Laboratory
    \and
    Christopher Watson \\
    University of Pennsylvania
    \and
    Junyu Zhou \\
    University of Pennsylvania
    \and
    Yuhao Liu \\
    University of Pennsylvania
    \and
    Chenxu Liu \\
    Pacific Northwest National Laboratory
    \and
    Ang Li \\
    Pacific Northwest National Laboratory
    \and
    Gushu Li \\
    University of Pennsylvania
    \and
    Samuel Stein \\
    Pacific Northwest National Laboratory
}

\begin{document}

\maketitle

\begin{abstract}

Compiling shallow and accurate quantum circuits for Hamiltonian simulation remains challenging due to hardware constraints and the combinatorial complexity of minimizing gate count and circuit depth. Existing optimization method pipelines rely on hand-engineered classical heuristics, which cannot learn input-dependent structure and therefore miss substantial opportunities for circuit reduction.

We introduce \textbf{F2}, an offline reinforcement learning framework that exploits free-fermionic structure to efficiently compile Trotter-based Hamiltonian simulation circuits. F2 provides (i) a reinforcement-learning environment over classically simulatable free-fermionic subroutines, (ii) architectural and objective-level inductive biases that stabilize long-horizon value learning, and (iii) a reversible synthetic-trajectory generation mechanism that consistently yields abundant, guaranteed-successful offline data.

Across benchmarks spanning lattice models, protein fragments, and crystalline materials (12-222 qubits), F2 reduces gate count by 47\% and depth by 38\% on average relative to strong baselines (Qiskit, Cirq/OpenFermion) while maintaining average errors of $10^{-7}$. These results show that aligning deep reinforcement learning with the algebraic structure of quantum dynamics enables substantial improvements in circuit synthesis, suggesting a promising direction for scalable, learning-based quantum compilation.

\end{abstract}

\section{Introduction}
\label{sec:intro}

Quantum computing offers a principled route to simulating quantum many‑body systems that are out of reach for classical resources, with anticipated impact on in~silico drug discovery, materials design and fundamental quantum chemistry research \cite{Feynman1982,Lloyd1996,AspuruGuzik2005,Peruzzo2014,Cao2019,McArdle2020, AspuruGuzik2005}. 
To model quantum systems of interest on quantum computers, the system’s dynamics must be expressed as machine-level instructions, commonly referred to as a quantum circuit. Such a circuit is typically constructed from a sequence of one- and two-qubit operations drawn from a fixed native gate set, which together approximate or exactly implement the target unitary evolution \cite{NielsenChuang2010,DawsonNielsen2006,RossSelinger2014,KliuchnikovMaslovMosca2013}. For instance, simulating the interaction between a protein and a small molecule requires representing its underlying quantum dynamics as such a circuit.

Near-term quantum architectures remain severely resource-limited: gate fidelities and restricted qubit connectivity constrain both circuit depth and total gate count \cite{Preskill2018}. As a result, aggressive circuit optimization is essential for achieving practical quantum advantage on near-term and emerging hardware. However, many core compilation objectives are computationally intractable; in particular, minimizing circuit depth or overall circuit complexity is NP-hard \cite{vanDeWeteringAmy2024,Ito2023Routing}.

A rich toolbox of compilers and rewrite systems exists, including industrial‑grade transpilers (e.g., Qiskit, t$|$ket$\,\rangle$, Cirq) and diagrammatic ZX‑calculus optimizers \cite{JavadiAbhari2024Qiskit,Sivarajah2020TKET,Duncan2020PyZX,CirqZenodo2024}. 
While deep learning has been applied to adjacent quantum-optimization tasks \cite{ruiz2024quantumcircuitoptimizationalphatensor, Tang2024AlphaRoute}, current neural networks struggle to learn general features of material and biochemical systems that enable efficient expressions as a quantum algorithm.

Three challenges in particular limit the applicability of deep learning to quantum program synthesis. First, quantum systems are widely believed to exhibit exponentially large feature spaces, making global generalization extremely difficult. Second, quantum-algorithm synthesis in an RL setting is a sparse-reward, long-horizon control problem, which complicates credit assignment. Third, gates combine discrete gate choices with continuous parameters, creating a large and heterogenous action space that is challenging to learn effectively.

We seek to address these challenges and make three contributions toward learning-based quantum program synthesis. \textbf{First}, we mitigate the exponential complexity of quantum-system representation and circuit optimization by introducing a reinforcement-learning environment tailored to classically simulatable subroutines, specifically free-fermionic subcircuits within larger quantum algorithms. \textbf{Second}, we develop a compositional action encoder together with a learned inductive bias in the critic objective, enabling more stable value estimation over a large, hybrid discrete-continuous action space. \textbf{Third}, we leverage the time-reversibility of quantum circuits to generate abundant synthetic trajectories; by sampling backward from the identity, we obtain guaranteed-successful transitions for offline reinforcement learning.

We evaluate F2 against state-of-the-art compilers, including IBM's Qiskit and Google's OpenFermion stack. On average, F2 achieves a 46\% reduction in gate count and a 36\% reduction in depth while maintaining approximation errors on the order of $10^{-7}$, demonstrating that learning aligned with physical structure can meaningfully improve quantum-circuit compilation.

\section{Background}

We provide here a brief overview of the concepts necessary for this work. For a comprehensive treatment of quantum computing fundamentals, we refer readers to \cite{nielsen_chuang_2010}.

\subsection{Hamiltonians and Time Evolution
\label{subsec:Hamiltonians}}

In quantum mechanics, a physical system is specified by its \emph{Hamiltonian} $\hat{H}$, a Hermitian operator. The time evolution of the system is given by the unitary operator
\begin{equation}
    U(t) = e^{-i \hat{H} t},
\end{equation}
which is obtained by exponentiating the Hamiltonian. Unitary operators satisfy $U^\dagger U = I$ and therefore preserve the normalization of quantum states, a fundamental requirement of physical evolution \cite{sakurai2014modern}. 
As $U(t)$ encodes both the energy spectrum and dynamical behavior of the system, simulating this evolution is central to many quantum-computing applications.

For many-body systems with $n$ particles, each having two degrees of freedom, the associated state space has dimension $2^n$. This exponential growth renders exact diagonalization of $\hat{H}$ or $U(t)$ intractable for all but the smallest systems. Quantum computers aim to circumvent this barrier by representing such unitaries efficiently as quantum circuits, motivating the development of algorithms that implement or approximate $U(t)$ on quantum hardware.

\subsection{Pauli Strings}
In an \(n\)-qubit system, a Pauli string is defined as a length-\(n\) tensor product of the Pauli gates \(\{X, Y, Z, I\}\), where each operator acts on a different qubit. This direct mapping of Pauli strings to qubits naturally arises in many quantum Hamiltonians, making them a convenient basis for both theoretical analyses and practical implementations.The time evolution generated by a single Pauli string, \(e^{-i P t}\), can be synthesized exactly using a sequence of Pauli gates, CNOT gates, and a single-qubit \(Z\)-rotation \cite{li2021paulihedralgeneralizedblockwisecompiler}.

Pauli strings of length \(n\) form a basis for the linear space of Hermitian 
operators on \(n\) qubits, and Hamiltonians are Hermitian. Consequently, any 
Hamiltonian can be decomposed as a weighted sum of Pauli strings, $H=\sum_iw_iP_i$ where $w_i\in\mathbb{R}$ \cite{li2021paulihedralgeneralizedblockwisecompiler}.

For notational simplicity, we absorb the coefficients into the operators and write \(H = \sum_i H_i\). While synthesizing the exponential of a single Pauli term \(e^{-i H_i t}\) is straightforward, synthesizing the exponential of a sum,\(\exp(it \sum_i H_i)\), is substantially more challenging. Closed-form circuit decompositions are generally not known, motivating approximation techniques that express the time evolution as a sequence of implementable operators.

A standard approach is \emph{Trotterization}, based on the Lie-Trotter formula \cite{Hatano_2005}

\begin{align}
e^{\,t\sum_{k=1}^{M} H_k}
&\approx
\left(
    \prod_{k=1}^{M}
        e^{\,\tfrac{t}{N} H_k}
\right)^{\!N},
\\[6pt]
\left\|
e^{\,t\sum_{k=1}^{M} H_k}
-
\left(
    \prod_{k=1}^{M}
        e^{\,\tfrac{t}{N} H_k}
\right)^{\!N}
\right\|
&\le
\frac{t^2}{2N}
\sum_{1 \le a < b \le M}
\|[H_a, H_b]\|
+\mathcal{O}\!\left(\frac{t^3}{N^2}\right)
\label{commutation_error}
\end{align}

where $t$ is the total evolution time and $N$ is the number of Trotter 
steps. The approximation error depends on the commutators \([H_a, H_b]\) and 
decreases linearly with \(N\). By decomposing a large Hamiltonian into 
smaller terms that can each be exponentiated exactly, Trotterization provides 
a systematic method for approximating time-evolution operators. Increasing 
the number of Trotter steps reduces the approximation error but increases 
circuit depth - a critical trade-off for near-term quantum devices.

Trotter-based methods are widely implemented in both industrial and academic 
software packages \cite{qiskit2024,cirq_developers_2024_11398048,Killoran_2019} 
and form the foundation of many Hamiltonian-simulation algorithms. The goal 
of our work is to deeply optimise this approximation process.

\subsection{Free Fermionic Systems}

Free fermionic, or quadratic, Hamiltonians form a class of quantum systems that can be simulated efficiently on classical hardware. We refer readers to \cite{Bravyi2005} for a comprehensive treatment. A \emph{mode} corresponds to a single fermionic degree of freedom, and fermions obey the Pauli exclusion principle, which prohibits multiple fermions from occupying the same mode. Creation and annihilation operators $(c_j^\dagger, c_j)$ respectively add or remove a fermion on mode $j$ and satisfy the canonical anti-commutation relations $\{c_j,c_k^\dagger\}=\delta_{jk}$ and $\{c_j,c_k\}=0$.

A general quadratic (or free) fermionic Hamiltonian takes the form
\begin{equation}
H=\sum_{j,k} A_{jk}\,c_j^\dagger c_k 
    + \tfrac{1}{2}\bigl(B_{jk}\,c_j^\dagger c_k^\dagger 
    + B_{jk}^*\,c_j c_k\bigr),
\end{equation}
The Heisenberg evolution of a time dependent operator $O(t)$ is governed by
\begin{equation}
\frac{dO(t)}{dt} = i[H, O(t)].
\end{equation}
For fermionic creation and annihilation operators, this equation closes linearly.

Let
\[
\mathbf{c}(t)
=
\begin{pmatrix}
c_1(t),\dots,c_n(t),c_1^\dagger(t),\dots,c_n^\dagger(t)
\end{pmatrix}^{\!\top},
\]
with the Heisenberg equation as a differential system
\begin{equation}
\frac{d\mathbf{c}(t)}{dt}
=
i\,[H,\,\mathbf{c}(t)]
=
K\,\mathbf{c}(t),
\label{eq:linear_eom}
\end{equation}
where $K$ is the $2n_f\times 2n_f$ matrix determined by the quadratic coefficients $(A,B)$ of $H$.

The solution of~\eqref{eq:linear_eom} is
\begin{equation}
\mathbf{c}(t)
= 
e^{Kt}\,\mathbf{c}(0)
=
U^\dagger(t)\,\mathbf{c}(0)\,U(t).
\end{equation}
Explicitly, the evolution takes the Bogoliubov–de Gennes form
\begin{equation}
\mathbf{c}(t)
=
\begin{pmatrix}
U_t & V_t\\[4pt]
V_t^* & U_t^*
\end{pmatrix}
\mathbf{c}(0),
\end{equation}
where $U_t$ and $V_t$ are $n_f\times n_f$ , and satisfy the canonical fermionic constraints \cite{Bravyi2005}.

Each operator at time $t$ is thus a linear combination of creation and annihilation operators across all modes:
\begin{equation}
c_j(t) = \sum_k \big(U_{t,jk}\,c_k + V_{t,jk}\,c_k^\dagger\big).
\end{equation}

This linear structure means that the full dynamical evolution is specified by the polynomial-sized matrices $(U_t, V_t)$, rather than by a state vector of dimension $2^n$. As a result, simulating free-fermionic dynamics requires only $O(n^2)$ parameters and is efficient to compute classically.

Physically, quadratic fermionic Hamiltonians describe systems whose dynamics are governed entirely by single-particle processes and pair creation or annihilation terms. Although modes may mix through $U_t$ and $V_t$, the absence of higher-order interactions prevents exponential complexity, making free-fermionic systems an important class of tractable quantum subsystems.

\paragraph{Mapping to Qubits.}
To implement these dynamics on quantum hardware, one applies a fermion–qubit mapping such as the Jordan–Wigner transformation,
\begin{equation}
c_j=\!\Big(\!\prod_{k<j}Z_k\!\Big)\frac{X_j-iY_j}{2},
\qquad
c_j^\dagger=\!\Big(\!\prod_{k<j}Z_k\!\Big)\frac{X_j+iY_j}{2}.
\end{equation}
This representation converts quadratic fermionic terms into tensor products of Pauli strings acting on neighboring qubits that obey fermionic statistics.  
Due to the representation of these fermionic operators in terms of Pauli strings, standard processes introduced in sections above allow for representations of free fermionic systems as quantum algorithms. 
Our research will be focused on finding these systems inside of interesting algorithms such as ones that simulate human proteins or highly correlated materials. 
Once these systems are found, our approach will be to express these subsystems as deeply optimized quantum subroutines of a target algorithm giving an optimization advantage.

\section{Reinforcement Learning Environment for \textsc{F2}
\label{sec:env}}

\subsection{Overview}
\begin{figure*}[t]
    \centering
    \includegraphics[width=\linewidth, height=0.42\textheight]{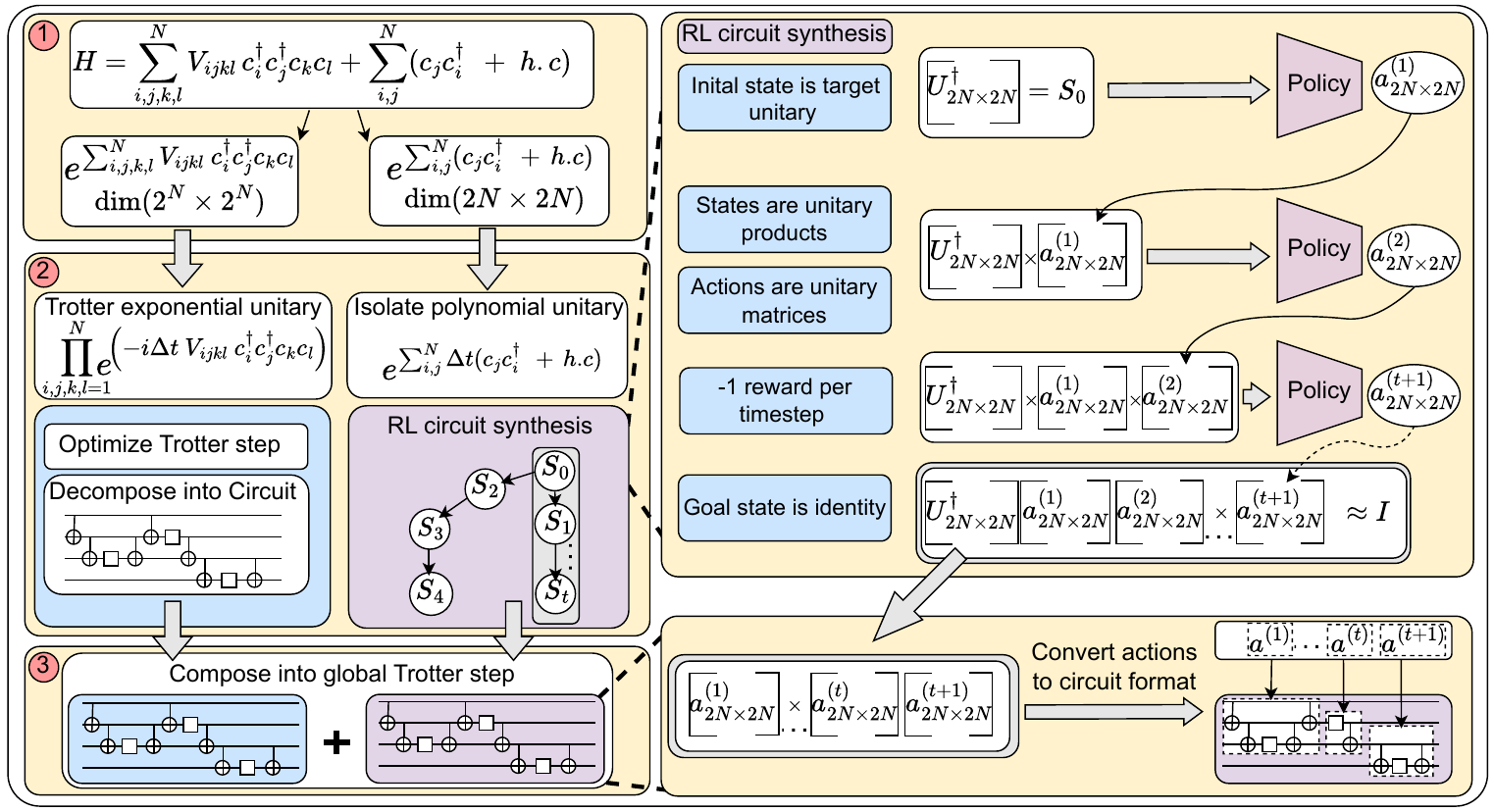}
    \caption{The goal is to optimize a Trotter step. 1) The input Hamiltonian is separated into classically representable terms and exponential terms. 2) Exponential unitaries are compiled using state-of-the-art techniques, while classically simulatable unitaries are decomposed with our reinforcement-learning algorithm. 3) The policy architecture constructs a trajectory by sequentially multiplying actions into the target unitary until the terminal state is reached.
    4) Actions have theoretical decompositions into circuit representations, creating a mapping between states and quantum circuits. 5) After both compilers complete circuit synthesis, the resulting circuits are concatenated to produce a deeply optimized Trotter step.}
    \label{fig:RLEnvironmentOverview}
\end{figure*}

Our agent targets depth and gate-efficient compilation of each Trotter step. 
The central observation is that many physically relevant quantum systems contain substructures that remain classically simulatable. 
Because Trotterization factorizes the full exponential into simpler component unitaries, these classically tractable pieces can be isolated and aggressively optimized (Fig. \ref{fig:RLEnvironmentOverview}). 
\textbf{Our method focuses on exploiting this structure: we identify the classically tractable factors within the Trotter step and optimize them via RL to improve the overall efficiency of the compiled evolution.}

\textbf{Scope and motivation.}
Here we will discuss how a reinforcement learning environment can be generated for the efficient compilation of free fermionic systems. 
We model the compilation of a free fermionic system as a Markov Decision Process with a discrete action space and deterministic dynamics. 
Unlike previous approaches to reinforcement learning for quantum algorithm synthesis, our environment restricts our model to walk along a Lie algebra of unitaries that have a guaranteed polynomial representation. 

Concretely, we expose the agent to a compact action alphabet built from exponentials of Pauli strings that map to these free‑fermionic evolutions via standard qubit–fermion encodings and gate decompositions.
We introduce the environment and dynamics for which the reinforcement learning agent will operate in and refer to Figure \ref{fig:RLEnvironmentOverview} for a visual overview.

\paragraph{State (residual unitary).}
Let $U^\star$ denote the target unitary we wish to synthesize. 
After $t$ time steps the agent has produced a partial synthesis $V_t=\prod_{k=1}^{t} A_k$, where each $A_k$ is one action (gate). 
We define the environment state at timestep $t$ as the product

\begin{equation}
S_t \;\coloneqq\; (U^\star)^\dagger V_t,
\end{equation}
so that $S_t \approx I$ if and only if the synthesized sequence approximately realizes $U^\star$. 
The agent observes this product matrix (which is polynomial in size due to our restriction of the action space), and the objective is to drive $S_t$ to the identity.

\paragraph{Action (Pauli‑string exponentials) and transition.}
Actions are elements of a discrete set of parameterized exponentials of Pauli strings drawn from a free‑fermionic generating set:

\begin{align}
\mathcal{A}
  &= \Big\{
      A(P,\theta)
      = \exp\!\big(-i\,\tfrac{\theta}{2}\,P\big)
      \;:\;
      P \in \mathcal{P},\;
      \theta \in \Theta
    \Big\},\\
  \mathcal{P} &= \big\{\, 
      X_i X_{i+1},\,
      Y_i Y_{i+1},\,
      X_i Y_{i+1},\,
      Y_i X_{i+1},\,
      Z_i
      \;|\;
      i = 1,\ldots,n-1
    \big\},\\
  \Theta &= \Big\{
      \pm \tfrac{\pi}{2^{k}}
      \;\big|\;
      k = 1,\ldots,20
    \Big\}
\end{align}

where $n$ is the number of qubits that $U^{\star}$ evolves.
A step applies the chosen gate $A_t$ to extend the synthesis $V_{t+1}=A_t V_t$, which updates the product to
\begin{equation}
S_{t+1}=(U^\star)^\dagger V_{t+1}=(U^\star)^\dagger A_t V_t.
\end{equation}
Since many gates do not commute, the ordering of actions materially affects $S_t$.

\paragraph{Reward, horizon, and termination.}
We use a sparse, length‑minimizing signal: each transition yields a reward $r_t=-1$, so the return equals minus the number of steps taken.
Episodes terminate on success or when a hard horizon is reached.
For our environment the fixed maximum episode length is $H_{max}=100$.
Success is detected by a unitary‑distance test against the identity, which can be expressed formally as: 
\begin{equation}
F_t \;=\;\frac{1}{d}\,\big|\mathrm{Tr}(S_t)\big|
\end{equation}
exceeds $1-\varepsilon$ (with tolerance $\varepsilon>0$).

Using the same notation, a transition in an episode at timestep $t$ can be represented as follows: 
\begin{equation}
\mathbf{f}_t \;=\; \bigl( S_t,\; A_t,\; r_t,\; S_{t+1}),
\label{eq:frame}
\end{equation}

\paragraph{Practical simulation and gate mapping.}
Although the environment is defined in full unitary form, all step dynamics are implemented in a polynomial‑time free‑fermion simulator. 

If $n_f$ is the number of fermions, our states will be represented by a $2n_f\times 2n_f$ real orthogonal matrix over fermionic modes. 
Here $n_f$ fermions are represented by $n=2n_f$ qubits. 
This makes \emph{step}, \emph{reset}, and \emph{check‑terminal} operations efficient. 
Each action $A(P,\theta)$ is compiled to a finite native gateset (e.g., CNOTs plus single‑qubit $R_z$) using standard decompositions, so the agent’s decisions lift directly to hardware‑compatible circuits for downstream applications in chemistry and materials.

\section{F2 Policy Architecture and Trajectory Generation}
\label{sec:f2}

We design an architecture that jointly encodes the \emph{quantum state}—represented as the product of unitaries already applied—and the \emph{action sequence} that produced it. 
These two views are complementary: the former captures the physical transformation in Hilbert space, while the latter records the syntactic structure of the synthesis process. 
The reason for both encoders is to have increased efficiency for learning in high variance labels due to Monte Carlo based rewards. 

To model both efficiently, we introduce \textbf{F2}, a dual-tower transformer composed of an \emph{axial-attention} \cite{ho2019axialattentionmultidimensionaltransformers} unitary encoder and a \emph{standard transformer} sequence encoder, fused into a lightweight value head (see Figure \ref{fig:architecture}). 
The two-tower pattern echoes AlphaTensor-style designs that pair structured tensor encoders with sequence models \cite{ruiz2024quantumcircuitoptimizationalphatensor}, while axial attention provides an efficient factorization of attention over multi-axis arrays \cite{ho2019axialattentionmultidimensionaltransformers}.

\subsection{Compositional Embedding}
\label{subsec:comp-embed}

Naïve one-hot action embeddings discard rich structure shared across operations (e.g., rotation axis, target qubits, angle magnitude). 
We instead use a novel \emph{compositional} representation that factorizes each action into semantically meaningful components which was co-designed with the environment. 
Concretely, write an action as
\[
a \;=\; (P,\theta), \qquad
P \in \{I,X,Y,Z\}^{\otimes n}\
\]
where \(P\) is a Pauli string identifying which operator acts on which qubit and \(\theta\) is the rotation parameter (discretized during training as in our environment). 
We embed each component separately—(i) Pauli string, (ii) Angle of rotation, (iii) the total length of the Pauli string, and (iv) the global index of the action for auxiliary features. 

Formally, the embedding can be expressed through the following function:
\begin{align}
\phi(a_i)
&=\; \text{TypeEmb}(\text{type}_i)
   + \text{AngleEmb}(\alpha_i) \\[3pt]
&\quad  
   + \text{WeightEmb}(|a_i|) +
   \text{GlobalEmb}(\text{idx}_i)
   \;\in\; \mathbb{R}^d .
\end{align}

Here, $d$ is the embedding vector dimension and all of the embedding functions are learned during training. 
This yields an embedding space in which actions that differ only by, say, angle or a single target index lie close to one another, providing the inductive bias needed to efficiently learn over large move sets with many overlapping features. 

\subsection{F2: A Dual-Tower Network for Quantum Circuit Synthesis}
\label{subsec:f2-arch}

\begin{figure}[t]
    \centering
    \includegraphics[width=\linewidth]{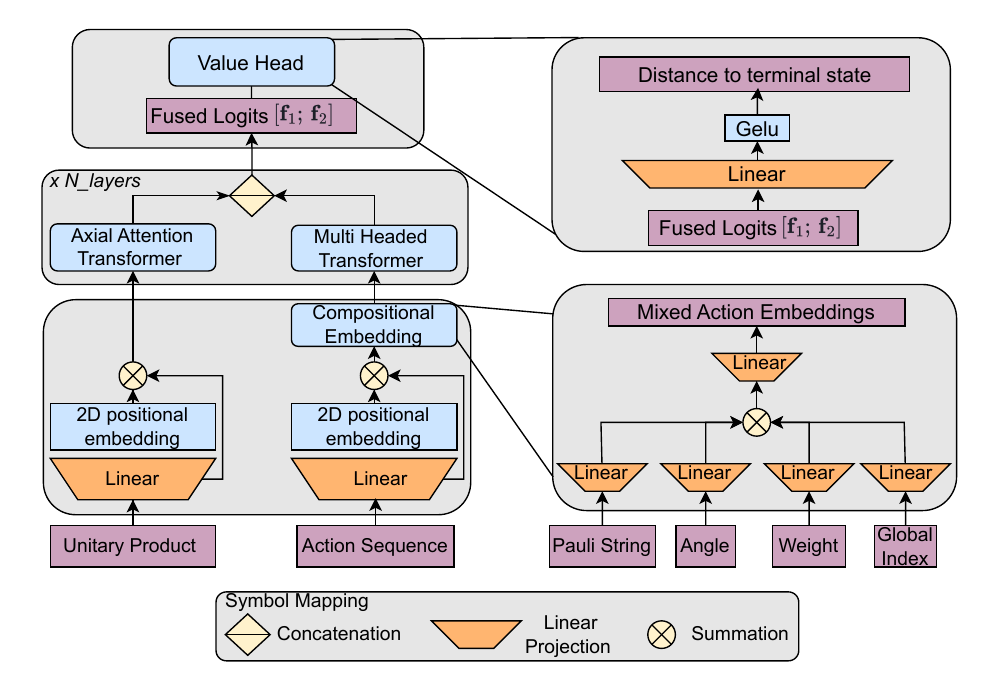}
    \caption{Our two tower neural network architecture using custom compositional embeddings. 
    The compositional embeddings factorize each feature of the action into a custom embedding and then mix these features together for the final embedding}
    \label{fig:architecture}
\end{figure}

Throughout this section, we will refer to Figure \ref{fig:architecture} to describe the architecture of our neural network. 

Let \(S_t\) denote the unitary product at time \(t\). 
We will refer to \(S_t\) as the residual unitary. 
We first project the entries of \(S_t\) to the model dimension and add two-dimensional positional encodings to preserve matrix geometry. 
We then apply \(N\) layers of \emph{axial attention} \cite{ho2019axialattentionmultidimensionaltransformers}, alternating attention along row and column axes to capture long-range algebraic interactions with tractable complexity. 
The result is a fixed-width representation \(h^{\text{unitary}}_t\).

Given the action history \(\{a_1,\dots,a_t\}\), we retrieve cached compositional embeddings for each \(a_k\) (Sec.~\ref{subsec:comp-embed}) and process the token sequence with a standard transformer stack. 
Mean pooling over the output tokens yields \(h^{\text{seq}}_t\).
Finally, we fuse towers via concatenation and a linear layer to form \(h^{\text{fused}}_t = \mathrm{Fuse}(h^{\text{unitary}}_t, h^{\text{seq}}_t)\).
A single-linear layer \emph{value head} predicts steps to terminal state from \(h^{\text{fused}}_t\).

\subsection{Trajectory-Reversal Data Augmentation}
\label{subsec:traj-rev}

Sparse rewards hamper exploration in long-horizon synthesis. 
We exploit the reversibility of quantum programs to generate \emph{guaranteed-successful} trajectories without environment interaction. 
Specifically, we \emph{walk backwards} from the identity unitary by sampling actions that (when inverted) define valid forward steps. 
Each frame \(s_t\) in a synthetic trajectory is labeled with the negative number of moves remaining to the terminal state.
These trajectories are then treated as off policy environmental interactions and used in offline reinforcement learning techniques for generating an efficient policy. 

Mathematically, we can express the trajectory reversal for data generation in the following way. 
In Section \ref{sec:env} our state is defined as: 
\begin{equation}
S_t \;\coloneqq\; (U^\star)^\dagger \prod_{k=1}^{t} A_k
\end{equation}

with the terminal goal state being defined by $S_t \approx I$. 
It is well known in quantum mechanics that a quantum unitary is reversible in time expressed as $UU^{\dagger} = I$. 
Here $U^{\dagger}$ is the time reversed unitary represented as a complex conjugate which inverts the evolution of $U$ from output state to input state.  
Given this time reversability, and the fact that all actions are unitary, our trajectories can be time reversed. 
\begin{equation}
U \;\coloneqq\; I \prod_{k=1}^{t} A_k   \Rightarrow
 U^{\dagger}\prod_{k=1}^{t} A_k=I
\end{equation}
When defining a unitary as $U$, our sequence of products now make a successful trajectory in our environment. 
By splitting each product at timestep $t$ into the frame 
\begin{equation}
\mathbf{f}_t \;=\; \bigl( (U^\star)^\dagger \prod_{k=1}^{t} A_k,\; A_{t+1},\; r_t,\; (U^\star)^\dagger \prod_{k=1}^{t+1} A_k),
\end{equation}
we can randomly sample actions to create successful trajectories. 
These trajectories can be regressed on to predict number of actions left in decomposition until identity matrix is reached. 

\subsection{Loss Function and Training
\label{subsec:objectivefunction}}
Here, we will introduce our training objective for the model to compile. 
The objective consists of two components: a Monte Carlo target which is the length of transitions left until termination from the current frame, and a geometric regularizer, which quantifies how close our state is to the terminal state. 

We will begin by discussing the main objective which is to learn the value function (distance from a state to the goal state in terms of actions).
Given our goal reaching synthetic trajectories, each frame is labeled with a negative scalar value $L - t=G_t$ where $L$ is the total sequence length and $t$ is the discrete timestep of the state. 
To fit our value head to this distance, Huber loss is used. 
Our first component of the objective is then $\mathcal{H}_\delta\!\big(G_t\big)$ where $\delta$ is the Huber loss parameter \cite{Huber_1964}. 

Now we will focus on the geometric regularizer of the loss function. 
Our regularizer is a scalar value denoted by $\Phi(S_t, \alpha, \beta)$ for our state $S_t$ and scalars $\alpha, \beta$. 
The goal is to add a learned inductive bias of the problem into the critic objective, which, to the best of our knowledge, has not been done before. 

Free fermionic states are able to be block diagonalized into a series of independent rotations. 
A special case is the terminal state where the block diagonalized form of the matrix has no rotations on independent subspaces. 
A geometric regularizer can then be fitted to the squared sum of the independent rotation angles where the goal state has the unique value of zero. 
To formalize, our state $S_t$ can be factorized as follows: 
\begin{equation}
\exists\,Q_t \in SO(2n_f)
\;\;:\;\;
Q_t^{\top}\,S_t\,Q_t
\;=\;
\bigoplus_{j=1}^{n_f}
\begin{bmatrix}
\cos\theta_{j} & \sin\theta_{j} \\[2pt]
-\sin\theta_{j} & \cos\theta_{j}
\end{bmatrix},
\end{equation}

Denoting $\phi(S_t) \;=\; \sum_{j=1}^{n_f} \theta_j^{\,2},$ then our geometric regularizer is defined as 
\begin{equation} 
\Phi(S_t, \alpha,\beta) = \alpha + \beta\phi(S_t)
\label{eq:thetasum}
\end{equation} 
Where the coefficients $\alpha, \beta$ are learned rather than set as hyperparameters to match scaling with our value function. 

Our total minimization objective is then formally expressed as: 
\begin{equation}
\mathcal{L}(\theta,\alpha,\beta)\;=
\mathcal{H}_\delta\!\big(G_t\big) + 
\Phi(S_t, \alpha,\beta)
\label{eqs:criticobjective}
\end{equation}

The parameters are learned such that our regularizer can properly match the magnitude of the Monte Carlo targets. 
The motivation behind this structure is so that values can have positive correlation with the geometric distance to the identity matrix which is predictive of how close our agent is to termination. 

The model is first fit to these targets using Monte Carlo based pretraining. 
Following Monte-Carlo pretraining for stability, we switch to TD updates to refine local consistency where in TD based training the geometric regularizer is dropped. 
The regimen of Monte Carlo pretraining prior to TD based learning is one that has been formally introduced in \cite{jelley2024efficientofflinereinforcementlearning}. 

Compared with generic time-symmetry or relabeling approaches for sparse-reward RL \cite{Barkley2024TimeSymRL, Andrychowicz2017HER, florensa2018reversecurriculumgenerationreinforcement, hoftmann2023backward, wang2021offline}, trajectory reversal leverages unitary invertibility to synthesize dense, on-goal successes that match our objective without goal relabeling.


\begin{table*}[t]
\centering
\fontsize{14pt}{18pt}\selectfont
\resizebox{\textwidth}{!}{%
\begin{tabular}{lcccccccc}
\hline
Benchmark (size) & G (F2) & D (F2) & G (Google-Cirq) & D (Google-Cirq) & G (IBM-base) & D (IBM-base) & G (IBM-rustiq) & D (IBM-rustiq) \\
\hline
2Cu2(CN)3  (32) & \textbf{8128} & \textbf{6481} & 13216 (38.50\%) & 7266 (10.80\%) & 16563 (50.93\%) & 9475 (31.60\%) & 14788 (45.04\%) & 13441 (51.78\%) \\
ABL Fragment (84) & \textbf{2591} & \textbf{1901} & 2696 (3.89\%) & 1906 (0.26\%) & 5048 (48.67\%) & 2827 (32.76\%) & 2800 (7.46\%) & 1904 (0.16\%) \\
Fermi Hubbard 1D (12) & \textbf{900} & \textbf{369} & 5100 (82.35\%) & 1809 (79.60\%) & 3110 (71.06\%) & 939 (60.70\%) & 2776 (67.58\%) & 2991 (87.66\%) \\
Fermi Hubbard 1D (72) & \textbf{624} & \textbf{110} & 3828 (83.70\%) & 381 (71.13\%) & 1580 (60.51\%) & 1026 (89.28\%) & 1762 (64.59\%) & 1713 (93.58\%) \\
Fermi Hubbard 1D (144) & \textbf{1272} & \textbf{110} & 7680 (83.44\%) & 381 (71.13\%) & 3200 (60.25\%) & 2070 (94.69\%) & 3526 (63.93\%) & 3477 (96.84\%) \\
Fermi Hubbard 2D (12) & \textbf{2560} & \textbf{2065} & 6760 (62.13\%) & 3711 (44.35\%) & 6235 (58.94\%) & 2860 (27.80\%) & 4177 (38.71\%) & 4983 (58.56\%) \\
Fermi Hubbard 2D (72) & \textbf{2710} & \textbf{1749} & 5914 (54.18\%) & 2028 (13.76\%) & 8926 (69.64\%) & 3780 (53.73\%) & 4729 (42.69\%) & 3497 (49.99\%) \\
Fermi Hubbard 2D (144) & \textbf{5114} & \textbf{2538} & 11522 (55.62\%) & 2841 (10.67\%) & 19696 (74.04\%) & 6963 (63.55\%) & 11423 (55.23\%) & 6957 (63.52\%) \\
Heisenberg 1D (12) & \textbf{1450} & \textbf{1011} & 6570 (77.93\%) & 3117 (67.56\%) & 2970 (51.18\%) & 1131 (10.61\%) & 2605 (44.34\%) & 3105 (67.44\%) \\
Heisenberg 1D (72) & \textbf{931} & \textbf{157} & 4003 (76.74\%) & 400 (60.75\%) & 1917 (51.43\%) & 2769 (94.33\%) & 997 (6.62\%) & 1439 (89.09\%) \\
Heisenberg 1D (144) & \textbf{1879} & \textbf{165} & 8023 (76.58\%) & 405 (59.26\%) & 3861 (51.33\%) & 5577 (97.04\%) & 1969 (4.57\%) & 2843 (94.20\%) \\
Heisenberg 2D (12) & \textbf{2490} & 3053 & 7610 (67.28\%) & 5643 (45.90\%) & 4590 (45.75\%) & \textbf{1833 (-66.56\%)} & 4156 (40.09\%) & 5172 (40.97\%) \\
Heisenberg 2D (72) & \textbf{2272} & \textbf{1408} & 5344 (57.49\%) & 1655 (14.92\%) & 3402 (33.22\%) & 1482 (4.99\%) & 2425 (6.31\%) & 2316 (39.21\%) \\
Heisenberg 2D (144) & \textbf{4998} & 2267 & 11142 (55.14\%) & 2505 (9.50\%) & 7128 (29.88\%) & \textbf{1716 (-32.11\%)} & 5566 (10.20\%) & 3673 (38.28\%) \\
La2CuO4 (16) & \textbf{6400} & 6096 & 18880 (66.10\%) & 10992 (44.54\%) & 10368 (38.27\%) & \textbf{2808 (-117.09\%)} & 9468 (32.40\%) & 11286 (45.99\%) \\
La2CuO4 (exp-fit U) (32) & \textbf{32960} & \textbf{22019} & 73664 (55.26\%) & 30819 (28.55\%) & 85232 (61.33\%) & 38127 (42.25\%) & 84727 (61.10\%) & 78103 (71.81\%) \\
La2CuO4 (theory U) (32) & \textbf{32960} & \textbf{22019} & 73664 (55.26\%) & 30819 (28.55\%) & 85232 (61.33\%) & 38127 (42.25\%) & 84725 (61.10\%) & 78104 (71.81\%) \\
PD-1 Fragment (28) & \textbf{279} & \textbf{167} & 348 (19.83\%) & 179 (6.70\%) & 581 (51.98\%) & 250 (33.20\%) & 321 (13.08\%) & 185 (9.73\%) \\
PD-1 Fragment + (74) & \textbf{3203} & \textbf{2182} & 4671 (31.43\%) & 2321 (5.99\%) & 5961 (46.27\%) & 2664 (18.09\%) & 3901 (17.89\%) & 4058 (46.23\%) \\
PD-1 Fragment ++ (222) & \textbf{24481} & \textbf{15899} & 30701 (20.26\%) & 16156 (1.59\%) & 43170 (43.29\%) & 22321 (28.77\%) & 45009 (45.61\%) & 33703 (52.83\%) \\
t-J 1D (12) & \textbf{3460} & \textbf{5314} & 7660 (54.83\%) & 6476 (17.94\%) & 12566 (72.47\%) & 7401 (28.20\%) & 4726 (26.79\%) & 6632 (19.87\%) \\
t-J 1D (72) & \textbf{2492} & \textbf{3571} & 5696 (56.25\%) & 3839 (6.98\%) & 7976 (68.76\%) & 10947 (67.38\%) & 2925 (14.80\%) & 4721 (24.36\%) \\
t-J 1D (144) & \textbf{5072} & \textbf{7197} & 11480 (55.82\%) & 7465 (3.59\%) & 16184 (68.66\%) & 22215 (67.60\%) & 5782 (12.28\%) & 9386 (23.32\%) \\
t-J 2D (12) & \textbf{5660} & \textbf{8122} & 9860 (42.60\%) & 9311 (12.77\%) & 19249 (70.60\%) & 17482 (53.54\%) & 6442 (12.14\%) & 8645 (6.05\%) \\
t-J 2D (72) & \textbf{7882} & \textbf{8013} & 11086 (28.90\%) & 8281 (3.24\%) & 26668 (70.44\%) & 23743 (66.25\%) & 8893 (11.37\%) & 9007 (11.04\%) \\
t-J 2D (144) & \textbf{22248} & \textbf{20232} & 28656 (22.36\%) & 20500 (1.31\%) & 73977 (69.93\%) & 69713 (70.98\%) & 29554 (24.72\%) & 26235 (22.88\%) \\
\hline
\end{tabular}
}
    \caption{Gatecount (G) and depth (D) comparisons across state-of-the-art compilation methods. 
    Next to each number is the reduction F2 has over that benchmark's performance. 
    The average reduction in gate count is 47\% and average reduction in depth is 38\% compared to our benchmarks.}\label{tab:gatecountDepthdata}
\end{table*}

\begin{figure*}[t]
    \centering
    \includegraphics[width=\linewidth]{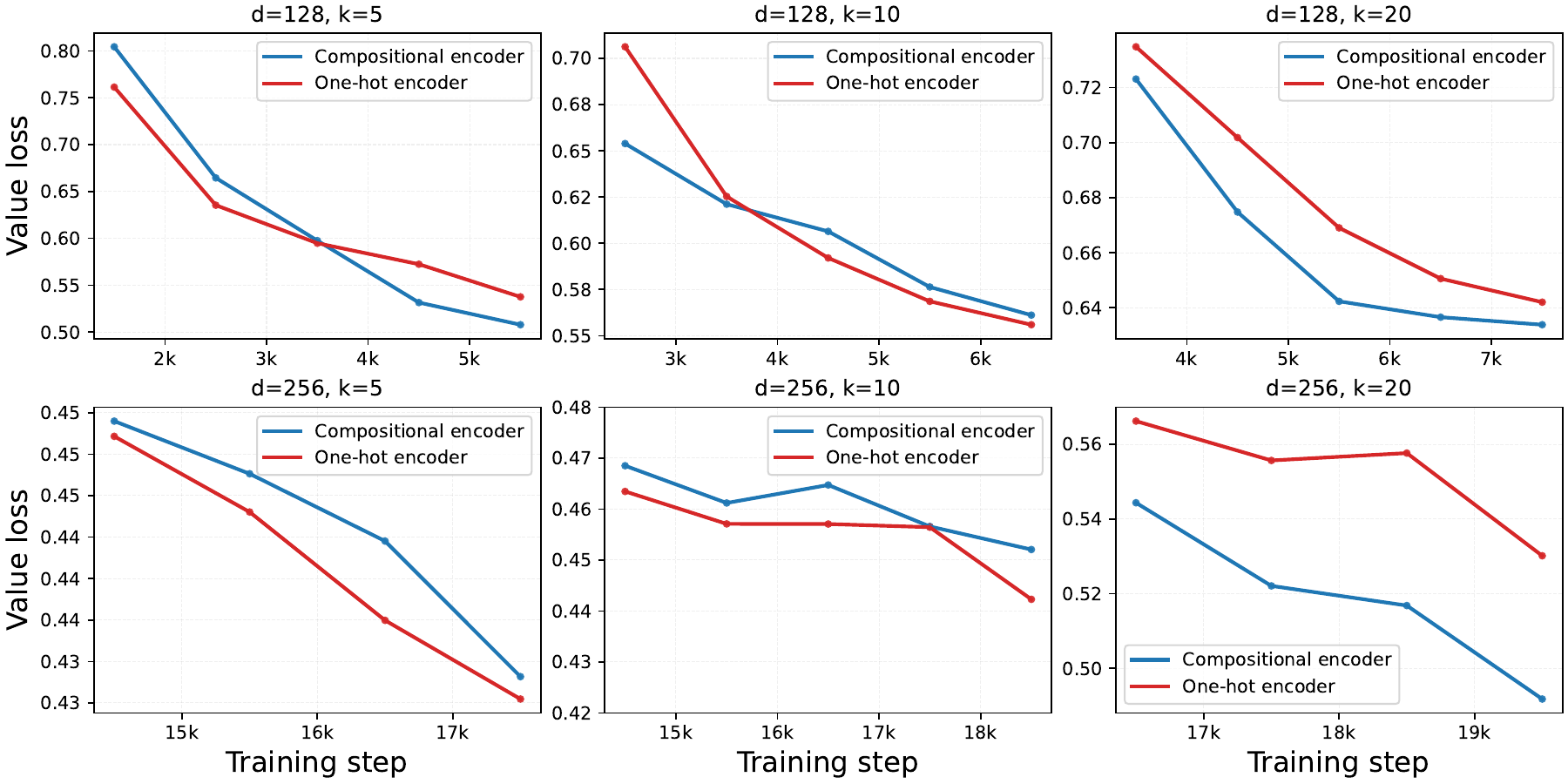}
    \caption{Comparing the learning efficency over one epoch on distance to identity regression. Architectures used are naive embeddings for action space vs compositional embeddings introduced in Figure \ref{fig:architecture}. X axis is on the order of 1000 steps}
    \label{fig:modelablation}
\end{figure*}

\section{Related Work}
\label{sec:related}
\paragraph{Trajectory data augmentation and generation.} Prior work on data augmentation and data generation for sparse-reward RL offers several complementary levers. Time-Symmetry Data Augmentation (TSDA) constructs reverse-time transitions by (approximate) time-reversibility without extra interaction, but the added samples are outcome-agnostic and do not transform failures into successes for a fixed goal \cite{Barkley2024TimeSymRL}. Hindsight Experience Replay (HER) relabels failed episodes with goals achieved along the trajectory to densify rewards; because it redefines the goal, it cannot create success for an unmet fixed goal \cite{Andrychowicz2017HER}. Reverse Curriculum Generation (RCG) adapts the start-state distribution by expanding outward from the goal and accepting starts whose empirical success lies in a target band, operating online under a reset-anywhere assumption \cite{florensa2018reversecurriculumgenerationreinforcement}. Offline RL lines of work train reverse (inverse-dynamics) world models to mitigate out-of-distribution error or improve data efficiency, but require offline datasets to fit the inverse dynamics \cite{hoftmann2023backward,wang2021offline}. In contrast, we keep the original goal fixed and exploit unitary-group structure together with a geometry-aware shaping signal to synthesize forward, successful rollouts from diverse initial states—expanding coverage near and far from the terminal manifold without goal relabeling or reverse-time samples.

\paragraph{Value function regularization.}


Prior work on augmenting value-function learning spans reward shaping, auxiliary tasks, and value-function regularization. 
Potential-based reward shaping~\cite{ng1999potential} provides a principled way to densify sparse rewards without altering the optimal policy, and typically acts directly on rewards through a manually tuned potential added to the extrinsic signal. 
Auxiliary-task methods such as \cite{jaderberg2016unreal,laskin2020curl,yarats2021drqv2,hafner2023dreamerv3} introduce additional learning objectives designed to build representations that support the downstream task, but these objectives are indirect and optimized via multiple heads that jointly shape a shared encoder. 
Value-function regularization approaches, commonly used in offline and conservative reinforcement learning \cite{kumar2020cql}, instead impose task-agnostic penalties to correct over-optimism on out-of-distribution states and stabilize value estimates.
Closest to our approach are approaches such as HuRL~\cite{cheng2021heuristicguidedreinforcementlearning} and HUBL~\cite{geng2023improving} that reduce variance and the effective decision making horizon by incorporating \textit{heuristics} (which ideally would estimate the optimal value function and may be e.g. hand engineered or Monte Carlo estimates from offline data)

In contrast to the approaches mentioned above, our method uses a single-headed architecture in which the value head's objective is modified by a geometric regularizer. 
Rather than adding a potential to the reward signal, we lift the heuristic into the value-loss function and learn the proper scaling in relation to extrinsic values.  
Unlike auxiliary-task formulations, our approach directly regulates the value estimate for the return of the state of interest.
Finally, unlike existing value-regularization methods that are task-agnostic and designed broadly for robustness, our formulation is explicitly optimized for generalization to off-policy states within the task itself.

\medskip
\noindent Deep learning architectures for quantum compilation have evolved from early single-qubit demonstrations to multi-qubit settings. 
\cite{Zhang2020TopologicalQuantumCompiling} presents one of the first RL-based approaches to quantum program synthesis but targets single-qubit unitaries only (whereas we scale to many qubit quantum algorithms) and does not consider efficient representations of full quantum algorithms or quantum \emph{systems}. 
Subsequent work extends to multi-qubit decompositions but typically assumes the system is already in circuit form and omits quantum-specific inductive biases \cite{Kolle2024RLQCS, Rietsch_2024}. \cite{ruiz2024quantumcircuitoptimizationalphatensor} introduces a dual-tower encoder over move histories and matrices for exact Clifford+T compilation, yet the architecture is specialized and does not leverage classically simulatable subroutines or shared structure across action subspaces. 
The newest iteration of RL unitary synthesis consists of online learning without special consideration for the unitary being compiled:  \cite{practicalRLsynthesis2024, zxcalcRL2025, directedQCS2024, paulinetworkRL2025}.

For routing, \cite{Tang2024AlphaRoute} applies a CNN to circuit routing with online RL and no synthetic data, again presuming a circuit-form input; a contemporaneous CNN-based routing approach appears in \cite{kremer2025optimizingnoncliffordcountunitarysynthesis}. 
Across these efforts, none exploit classically simulatable kernels of quantum systems to obtain efficiently sized state representations.

\medskip
\noindent Beyond RL, non–AI compiler techniques for Hamiltonian simulation—which is our focus—have progressed rapidly. 
Early efforts relied on simultaneous diagonalization of commuting Pauli strings \cite{Cowtan_2020, van_den_Berg_2020}. 
These are later outperformed by reordering-driven gate-cancellation strategies \cite{li2021paulihedralgeneralizedblockwisecompiler, Gui_2020_TermGrouping, Anastasiou_2022_TETRIS_ADAPT_VQE} and by Pauli-network synthesis methods \cite{deBrugiere_2024_ShorterSynthesis,Paykin_2023_PCOAST, Schmitz_2024_GraphOptimizationPerspective, decker2025kernpilercompileroptimizationquantum} that substantially reduce the depth of Trotter steps. 
The QuCLEAR framework \cite{Liu_2024_QuCLEAR} further explores Clifford extraction and absorption for Hamiltonian simulation, but requires updating the observable; importantly, it leaves other circuit components untouched, so the compiled time-evolution operator remains reusable. 
Other approaches propose grouping non-commuting terms in Trotter steps in order to increase overall accuracy of the approximation method, but do not find significant optimizations over state of the art for per Trotter-step gate count \cite{decker2025kernpilercompileroptimizationquantum}.
Finally, mapping-aware optimizations such as Fermihedral and HATT target fermion-to-qubit encodings to cancel gates \cite{Liu_2024, Liu_2025}, though these benefits are specific to Fermionic Hamiltonians.

\section{Experiments}
\label{sec:experiments}
Here we discuss the experimental setup and evaluation of our work. 

\subsection{Evaluation}
\label{subsec:evaluation}


\begin{figure}[t]
    \centering
    \includegraphics[width=\linewidth]{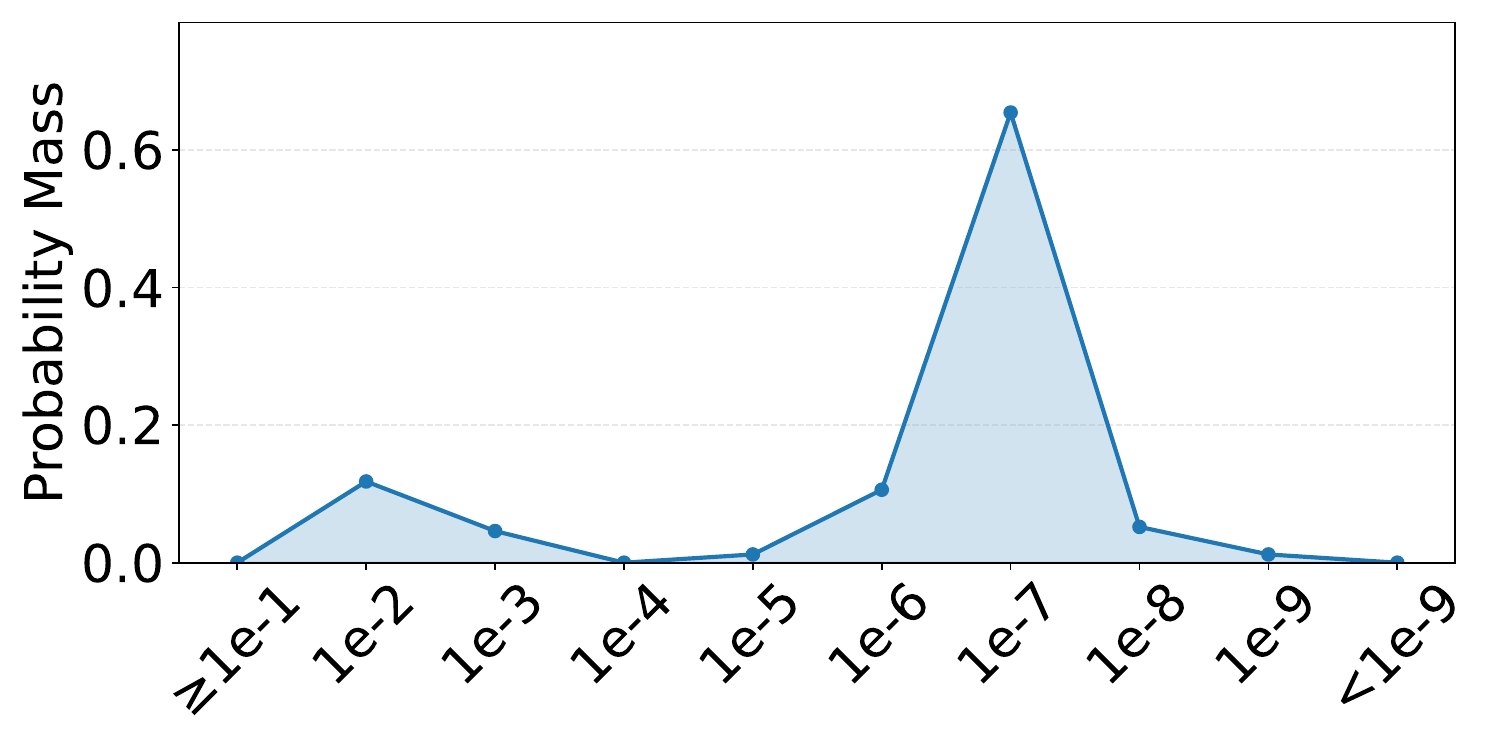}
    \caption{500 random unitaries with small angle rotations compiled by F2. Fidelity is grouped by error ($1-F$).}
    \label{fig:fidelitybarchart}
\end{figure}

\paragraph{Experimental setup.}
Training was performed on a single NVIDIA A100 GPU and an AMD EPYC~9654P 96-core CPU. 
Our implementation uses PyTorch~2.5.1 with CUDA~12.1. 
We compare against public, stable baselines: Qiskit~1.3.2 \emph{(optimization level 3 for all IBM runs)}, OpenFermion~1.7.0, and Cirq~1.5.0 \cite{JavadiAbhari2024Qiskit,cirq_developers_2024_11398048}. 
Google-stack circuits were compiled through Cirq (not passed through IBM’s optimizer). 
F2 outputs are emitted directly as hardware-compatible circuits via our custom implementation without routing through IBM’s transpiler.

\paragraph{Benchmarks.}
We evaluate our method across three domains spanning both foundational science and commercially relevant workloads: (i) strongly correlated physics models, (ii) crystalline materials, and (iii) protein fragments. 
First, we include the Fermi--Hubbard, Heisenberg, and $t$--$J$ models, which arise as effective electronic Hamiltonians in appropriate limits and are widely used to study quantum phases such as spin liquids and quantum glasses \cite{Hubbard1963, Anderson1959, Chao1977}. 
Second, we evaluate on two experimentally important materials---La$_2$CuO$_4$ and $\kappa$-(ET)$_2$Cu$_2$(CN)$_3$---whose low-energy behavior can be mapped onto Hubbard- or Heisenberg-type models with parameters obtained from \emph{ab initio} and fitted calculations \cite{Lane2018,Zheng2005,Coldea2001, Nakamura2009}. 
Finally, we include fragments of biologically relevant proteins. 
Modeling full proteins requires thousands of qubits, but focusing on functional hotspots yields tractable subsystems while preserving biochemical relevance \cite{Berman2000}. 
We use fragments from PD-1 and ABL1, two human membrane-associated proteins involved in immune regulation and cell-cycle control, and key targets in cancer therapeutics. 
Across all domains, problem sizes range from $12$ to $222$ qubits and were implemented for compilation using OpenFermion with our neural network having a maximum error allowance of $10^{-6}$.

\paragraph{Metrics.}  
We evaluate compiler performance using three quantitative metrics. 
The \textbf{circuit depth} \(D\) measures the  (parallelism) depth of the circuit after applying standard commutation and gate-merging passes, reflecting temporal efficiency. 
The \textbf{gatecount}  \(G\) represents the total number of quantum operations on the computer. 
Finally, we compute the \textbf{fidelity} with respect to the target unitary \(U_{\star}\) as
\[
F := \Bigl|\tfrac{1}{d}\operatorname{Tr}(U_{\star}^{\dagger}\,\hat U)\Bigr|,
\]
where \(d\) is the Hilbert-space dimension and \(\hat U\) is the compiled unitary.

\paragraph{Depth and gate count.}
Table~\ref{tab:gatecountDepthdata} summarizes depth and gatecount across the suite. 
On average, our agent can reduce gatecounts by 47\% and depth by 38\% respectively with maximum reductions of 83\% and 97\%. 
While F2 achieves consistent reductions versus industry standard compilers, the largest reductions observed are  on materials and fundamental-model families. 

We attribute these gains to parameter regularity in the benchmark, which concentrates input angles into narrow bands and makes our compositional action space easier to exploit. 
In contrast, molecular Hamiltonians exhibit broader coefficient distributions; this stresses angle discretizations and increases the likelihood that the agent must factor more extreme inputs into the move set, modestly attenuating improvements.

\paragraph{Parallelism.}
Improvements in depth are positive but smaller than gate count reductions. 
Two factors contribute: (i) the reward function penalizes long sequences but does not finely differentiate operations with very different downstream parallel structure, and (ii) discretizing continuous angles can initially inflate depth until standard merging/commutation cleans it up.

Let's start with a concrete example for why our first factor can contribute to less optimal depth. 
Consider two actions from our set $XX(\theta)$ and $Z(\theta)$. 
Both actions will accumulate a reward of $r_t = -1$ however $XX$ will require many more operations than $Z$ in circuit representation due to the analytical decomposition for each. 

Second, due to the choice of discretizing angles, our model must create longer sequences than necessary which must be post processed. 
The reason for long sequences is that the continuous parameters must be factorized into a sum of discrete angles each one time step long.
Post-pass optimization recovers a portion of this gap, but we leave parallelism-aware shaping and more finegrained reward functions for future work.

\paragraph{Approximate synthesis and accuracy.}
Our approach targets \emph{approximate} unitaries within a prescribed tolerance rather than analytical identities. 
Given hardware noise, many applications only require bounded error (e.g., domain-specific “chemical accuracy” thresholds \cite{Louvet2023Feasibility}) rather than exact synthesis. 
This flexibility expands the optimization surface: F2 can trade negligible accuracy for sizable circuit savings. 
Figure~\ref{fig:fidelitybarchart} reports compilation fidelity across random unitaries found in Trotter steps; accuracy remains within target tolerances while improving \(D\) and gatecount.
Here, Hamiltonians and coefficients are sampled from our benchmark with the range of $\Delta t$ between $(-0.02, 0.02)$, which is commonly cited as the target Trotter step size interval \cite{Heyl2019TrotterLocalization, Babbush2014ChemTrotterError, Yi2022SpectralPF, Poulin2015TrotterChem}. 
For our work and the current trajectory of quantum computing, an error of $10^{-7}$ is justified for many near and intermediate-term Hamiltonian simulations without compromising accuracy targets. 

\paragraph{Action Embedding Evaluation}
Here we show the efficacy of our custom action embedding compared to a naive embedding architecture over the action space. 

Our experiment compares loss curves for predicting distance to terminal state based on Monte Carlo based labels and comparisons were done for varying sized action sets and varying feature space dimensions. 
Results of this experiment are captured in Figure \ref{fig:modelablation}. 

What is noticed is that as our action space is increasing, there is more learning efficiency observed our proposed architecture. 
This is optimistic when scaling our neural network to handle even larger unitaries as gains are already seen at a small scale.

\begin{figure}[t]
    \centering
    \includegraphics[width=\linewidth]{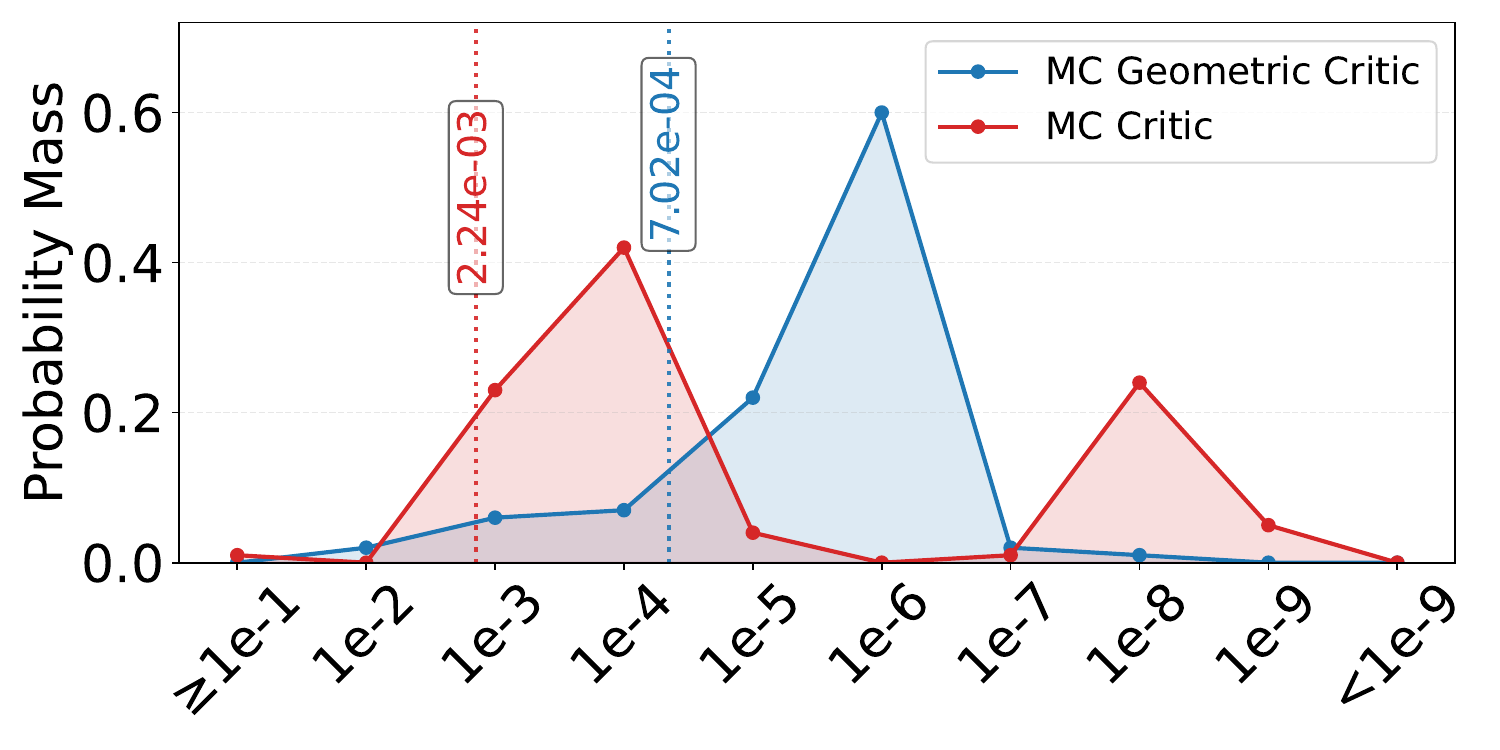}
    \caption{100 random unitaries with small angle rotations compiled by Monte Carlo pretrained F2. 
    Geometric critic represents our regularized objective with learned weightings while critic is a model trained on the reward function shaped by geometric returns at a flat scaling. 
    Fidelity is grouped by error ($1-F$).}
    \label{fig:geometricablation}
\end{figure}

\paragraph{Critic Geometric Objective Evaluation}
Now the performance changes due to the learned regularizer of equation \ref{eqs:criticobjective}. 
The evaluation compared our critic objective to one where the learned geometric heuristic was dropped and instead the reward was shaped by a potential reward shaping function: 
\begin{equation}
r_t = -1  + \beta\bigl(\phi(S_{t+1}) - \phi(S_t)\bigr)
\end{equation}
where $\phi(S_t)$ is defined in equation \ref{eq:thetasum}.
In our environment, future rewards are not discounted so there is no discount factor and $\beta$ was set to 2. 
In the experiment models were prepared by training on the same training set to the same number of epoch iterations. 
The rewards regressed on were Monte Carlo rewards only and we did not perform a TD-based fine-tuning step afterwards. 
Next, over 100 random unitaries, the models would perform compilation and the approximation accuracy would be recorded. 
The approximation error distributions along with the means are plotted in Figure \ref{fig:geometricablation}. 
As noticed, our pretrained models gain a decreased approximation error by an order of magnitude compared to the canonical potential-based reward shaping. 

The explanation for this result follows from two features of the implementation. Potential-based rewards are typically used to densify sparse feedback and signal that states closer to the goal are more valuable. In our setting, however, every trajectory is guaranteed to reach the goal, and trajectories are sampled without regard to geometric proximity. As a consequence, a state physically near the goal may still receive a large “distance-to-goal” label. This randomness makes it extremely difficult to hand-tune the scaling of a potential-based shaping term. A simpler alternative is to recognize the positive correlation between geometric distance and distance-to-goal labels and to learn the appropriate scaling factor directly. While this changes the optimal policy relative to classical potential-based shaping, it greatly reduces the amount of error that can enter the learning process through manual hyperparameter tuning in an offline setting.

\section{Conclusion}
\label{sec:conclusion}
In conclusion, this work has shown that offline reinforcement learning can be used for efficient compilation of quantum algorithms. 
Our work has shown gate count and depth reductions on average above 36\% with maximum reductions of over 80\% compared to baselines considered. 

While this progress is promising, multiple research questions are still unanswered. 
These questions are as follows. 
How could this paradigm be extended to other classically tractable subroutines, such as those efficiently tractable by tensor networks? 
Can the same performance be achieved or improved upon through the incorporation of continuous parameters into the move space? 
Can architectural characteristics be considered by the model? 

Many more questions exist for how AI techniques can be specialized such as paradigms that use both offline and online reinforcement learning. 
Overall, we believe these questions are important with answers that can bring an optimistic future to quantum computing. 


\section*{Acknowledgements}
This research was supported by PNNL’s Quantum Algorithms and Architecture for Domain Science (QuAADS) Laboratory Directed Research and Development (LDRD) Initiative. This material is based upon work supported by the U.S. Department of Energy, Office of Science, National Quantum Information Science Research Centers, Quantum Science Center (QSC). The Pacific Northwest National Laboratory is operated by Battelle for the U.S. Department of Energy under Contract DE-AC05-76RL01830. This research used resources of the Oak Ridge Leadership Computing Facility (OLCF), which is a DOE Office of Science User Facility supported under Contract DE-AC05-00OR22725. This research used resources of the National Energy Research Scientific Computing Center (NERSC), a U.S. Department of Energy Office of Science User Facility located at Lawrence Berkeley National Laboratory, operated under Contract No. DE-AC02-05CH11231. GL, JZ, and YL were in part supported by the U.S. Department
of Energy, Office of Science, Office of Advanced Scientific
Computing Research through the Accelerated Research in
Quantum Computing Program MACH-Q project., NSF CAREER Award No. CCF-2338773 and ExpandQISE Award No. OSI-2427020.


\bibliographystyle{IEEEtranS}  
\bibliography{example_paper}

\end{document}